\documentclass[%
 reprint,
superscriptaddress,
 amsmath,amssymb,
 aps
]{revtex4-2}

\usepackage[pdftex, pdftitle={Article}, pdfauthor={Author}]{hyperref} 
\usepackage{graphicx}
\usepackage{dcolumn}
\usepackage{bm}

\begin{document}

\preprint{APS/123-QED}

\title{Basis Function Dependence of Estimation Precision \\for Synchrotron-Radiation-Based Mössbauer Spectroscopy}
\author{Binsheu Shieh}%
   \affiliation{%
     Graduate School of Frontier Sciences, The University of Tokyo, Kashiwa, Chiba 277-8561, Japan
    }%

\author{Ryo Masuda}%
    \affiliation{%
        Graduate School of Science and Technology, Hirosaki University, Hirosaki, Aomori 036-8561, Japan
    }%

\author{Satoshi Tsutsui}%
    \affiliation{%
     Japan Synchrotron Radiation Research Institute (JASRI), Sayo, Hyogo 679-5198, Japan}%
    \affiliation{%
     Institute of Quantum Beam Science, Graduate School of Science and Engineering, Ibaraki University}%
    
\author{Shun Katakami}%
    \affiliation{%
        Graduate School of Frontier Sciences, The University of Tokyo, Kashiwa, Chiba 277-8561, Japan
    }%
\author{Kenji Nagata}%
    \affiliation{%
        Center for Basic Research on Materials (CBRM), National Institute for Materials Science (NIMS), Tsukuba, Ibaraki 305-0044, Japan
    }%
\author{Masaichiro Mizumaki}%
    \affiliation{%
            Faculty of Advanced Science and Technology, Kumamoto University, Kumamoto, Kumamoto 860-8555, Japan
        }%

\author{Masato Okada}
    \affiliation{%
        Graduate School of Frontier Sciences, The University of Tokyo, Kashiwa, Chiba 277-8561, Japan
    }%


\date{\today} 

\date{\today}

\begin{abstract}
Mössbauer spectroscopy is a technique employed to investigate the microscopic properties of materials using transitions between energy levels in the nuclei. Conventionally, in synchrotron-radiation-based Mössbauer spectroscopy, the measurement window is decided by the researcher heuristically, although this decision has a significant impact on the shape of the measurement spectra. In this paper, we propose a method for evaluating the precision of the spectral position by introducing Bayesian estimation. The proposed method makes it possible to select the best measurement window by calculating the precision of Mössbauer spectroscopy from the data. Based on the results, the precision of the Mössbauer center shifts improved by more than three times compared with the results achieved with the conventional simple fitting method using the Lorentzian function.
\end{abstract}

\maketitle


\section{Introduction}
Mössbauer spectroscopy is an technique employed to investigate the properties of materials based on the nuclear transitions occurring between specific energy levels in the nuclei ~\cite{zhdanov1968crystal}. This technique enables the study of the hyperfine interactions and vibrational states of atoms at the probe nucleus. Therefore, Mössbauer spectroscopy has been applied in the fields of materials science, solid-state physics, geoscience, and biophysics. Accordingly, this technique has contributed to the elucidation of the functions and properties of new materials by revealing insights into their microscopic electronic states and atomic dynamics~\cite{guetlich2011}, ~\cite{seto2015}.

Mössbauer spectroscopy using radioactive isotopes (RI Mössbauer spectroscopy) is a conventional Mössbauer spectroscopy technique that employs radioactive isotopes (RIs) as the beam source. In RI Mössbauer spectroscopy, the sample to be measured is irradiated with gamma rays to investigate the hyperfine interactions at the probe nucleus position in the material ~\cite{zhdanov1968crystal}. Synchrotron-radiation-based Mössbauer spectroscopy (SR-based Mössbauer spectroscopy) adopts SR as the incident radiation instead of gamma rays. This technique proceeds as follows ~\cite{seto2009synchrotron},~\cite{seto2010mossbauer},~\cite{seto2021synchrotron}: two samples containing the same resonant Mössbauer nuclei are prepared, one as a transmitter and the other as a scatterer. The experimental setup includes monochromatized SR and a silicon Avalanche photodiode detector to measure the Mössbauer spectra, which reflect the hyperfine interactions at the probe nucleus position. The probe nuclei in the sample excited by the SR release internal conversion electrons ~\cite{masuda2014synchrotron}, fluorescent X-rays, and gamma rays during de-excitation. When the resonance energies of the scatterer and the transmitter coincide, the SR absorption by the transmitter is enhanced and the scattering from the scatterer is reduced. The resonance energy of the transmitter or the scatterer is controlled by the Doppler effect of light, i.e., by varying its velocity. Thus, by measuring the scattering from the scatterer as a function of the energy difference between the resonance energy of the scatterer and that of the transmitter, a local minimum (absorption peak in the spectrum) can be observed when the energy difference is zero. The scattering intensity that depends on this energy difference is graphed as the Mössbauer spectrum.

Theoretically, the SR-based Mössbauer spectrum is determined by two factors. The first is the hyperfine interaction, which is the interaction between electrons and the nucleus. The hyperfine interaction reflects the electron density, electric field gradient, and magnetic fields created by electrons at the nucleus. The second is the resolution function of the Mössbauer spectroscopy measurement system. This is a unique property brought about by the use of pulsed SR, and it is not usually treated in RI Mössbauer spectroscopy where nuclear gamma rays are employed instead. In this study, we focus on the shape of the resolution function. The spectral resolution function in SR-based Mössbauer spectroscopy is determined by two parameters:  $\tau_1$ and $\tau_2$ ($ \tau_1<\tau_2$), which represent the start and end of the measurement window ~\cite{seto2009synchrotron}, ~\cite{seto2010mossbauer}. Note that when $\tau_1=0$ and $\tau_2=\infty$, the function is approximated by the Lorentzian function used in the simple SR-based Mössbauer spectroscopy ~\cite{seto2012condensed},~\cite{matsuoka2011structural},~\cite{tsutsui2018precise}. Furthermore, the shape of the resolution function depends on $\tau_1$ and $\tau_2$. Although the measurement window, determined by $\tau_1$ and $\tau_2$, affects the counting efficiency and energy resolution, which are in the relation of trade-off for the precise determination of the absorption peak, the determination of this parameter is left to experts. For a more precise analysis of the spectra, it is necessary to establish a theoretical basis for selecting the optimum measurement window.

Recently, Bayesian estimation has found application in various areas of solid-state physics~\cite{Akai_2018},~\cite{Mototake2019},~\cite{Kashiwamura2022},~\cite{Ueda2023}, including Mössbauer spectroscopy ~\cite{moriguchi2022bayesian}. In this study, to overcome the aforementioned problems, we propose a new numerical method that uses Bayesian estimation to analyze the peak positions in the SR-based Mössbauer spectroscopy spectra. We introduce a discretization calculation method to efficiently compute multiple integrals in the resolution function. The Bayesian estimation approach makes it possible to obtain information about the absorption peak position in the spectrum as a probability distribution from the resolution function and the measurement window obtained from past experiments ~\cite{RevModPhys.83.943},~\cite{nagata2012bayesian},~\cite{Tokuda2017}, ~\cite{nagata2019bayesian}. Our proposed method allows the discussion of the resolution function position and a comparison of different measurement windows. We show that our proposed method improves the estimation precision of the resolution function based on the measurement window, compared with the precision achieved using the conventional simple method, i.e., by fitting with the Lorentzian function. Here, precision is defined as the degree to which the peak position estimated from the data coincides with the true peak position.

The paper is organized as follows. In Section \ref{sec:formulation}, we introduce the resolution function and the Bayesian estimation framework for SR-based Mössbauer spectroscopy. In Section \ref{sec:numerical}, we evaluate the effectiveness of the proposed method by analyzing the generated data. In Section \ref{sec:summary}, we employ the analysis results to discuss the measurement window. Section 5 presents the conclusions.

\section{Formulation}\label{sec:formulation}

In this section, we describe the spectral resolution function of the measurement system for the SR-based Mössbauer spectroscopy and a framework for introducing Bayesian estimation into the resolution function. A method for estimating the center position of the absorption peak in an SR-based Mössbauer spectroscopy spectrum by Bayesian estimation is also presented.

\subsection{Model}\label{sec:model}
Here, we introduce the spectral resolution function of the SR-based Mössbauer spectroscopy measurement system based on the information in \cite{seto2010mossbauer}. We simulated the resolution function using 151Eu Mössbauer spectroscopy with 21.5 keV resonance. This was applied to estimate the absorption peak in the Mössbauer spectrum using Bayesian estimation. A parameter that indicates the center position of the absorption peak of the spectrum is introduced in the resolution function.

\begin{widetext}
    \begin{equation}
        I\left(w_s,\Delta w, \tau_1, \tau_2\right)=I_{\mathrm{A}}\left(w_s,\Delta w, \tau_1, \tau_2\right)+I_{\mathrm{C}}\left(w_s,\Delta w, \tau_1, \tau_2\right)+I_{\mathrm{B}} 
    \end{equation}
    \begin{equation}
        I_{\mathrm{A}}\left(w_s,\Delta w, \tau_1, \tau_2\right)=C_{\mathrm{A}} \int_{\tau_1}^{\tau_2} \mathrm{~d} \tau \int_0^{z_{\mathrm{s}}} \mathrm{d} z \left|\int \frac{\mathrm{d} w}{2 \pi} \frac{\exp (-\mathrm{i} w \tau)}{w-(w_{\mathrm{s}}+\Delta w)+\mathrm{i} / 2} E_{\mathrm{t}}(w) E_{\mathrm{s}}\left(w, w_{\mathrm{s}}+\Delta w, z\right)\right|^2
    \end{equation}
    \begin{equation}
        I_{\mathrm{C}}\left(w_s,\Delta w, \tau_1, \tau_2\right)=C_{\mathrm{C}} \int_{\tau_1}^{\tau_2} \mathrm{~d} \tau \int_0^{z_s} \mathrm{~d} z \left|\int \frac{\mathrm{d} w}{2 \pi} \exp (-\mathrm{i} w \tau)\left(E_{\mathrm{t}}(w) E_{\mathrm{s}}\left(w, w_{\mathrm{s}}+\Delta w, z\right)-1\right)\right|^2 
    \end{equation}
    \begin{equation}
        E_{\mathrm{t}}(w)=E_{0 \mathrm{t}} \exp \left(-\frac{\mu_{\mathrm{et}} z_{\mathrm{t}}}{2}\right) \exp \left(-\mathrm{i} \frac{\mu_{\mathrm{nt}} z_{\mathrm{t}}}{2(2 w+i)}\right) 
    \end{equation}
    \begin{equation}
        E_{\mathrm{s}}\left(w, w_s, z\right)=E_{0\mathrm{s}} \exp \left(-\frac{\mu_{\mathrm{es}} z}{2}\right) \exp \left(-\mathrm{i} \frac{\mu_{\mathrm{ns}} z}{2\left(2\left(w-w_{\mathrm{s}}\right)+\mathrm{i}\right)}\right)
    \end{equation}
\end{widetext}

The SR-based Mössbauer spectra are represented by $I\left(w_s,\Delta w,\tau_1,\tau_2\right)$ in Eq. (1), where $w_s$ is the energy difference between the transmitter and the scatterer. As shown in Eq. (1), $I\left(w_s,\Delta w, \tau_1, \tau_2\right)$ is the sum of three channels, $I_A\left(w_s,\Delta w, \tau_1, \tau_2\right)$, $I_C\left(w_s,\Delta w, \tau_1, \tau_2\right)$, and $I_B$. $I_A\left(w_s,\Delta w, \tau_1, \tau_2\right)$ represents the nuclear resonant scattering of SR without recoil at the scatterer, $I_C\left(w_s,\Delta w, \tau_1, \tau_2\right)$ represents the scattering by the photoelectric effect at the scatterer, $I_B$ represents other scattering processes, including nuclear resonant inelastic scattering. Details of the calculations of $I_A\left(w_s,\Delta w, \tau_1, \tau_2\right)$ and $I_C\left(w_s,\Delta w, \tau_1, \tau_2\right)$ are expressed in Eq. (2)-Eq. (5), where $z_t(z_s)$ is the thickness of the transmitters (scatterers), $\mu_{nt}(\mu_{ns})$ is the linear absorption coefficient for nuclear resonance in the transmitters (scatterers), $\mu_{et}(\mu_{es})$ is the electronic absorption coefficient of the transmitters (scatterers), $E_{t}(E_{s})$ is the amplitude of the propagating coherent radiation field of the transmitter (scatterer), and $E_{0t}(E_{0s})$ is the amplitude of the radiation field at the incidence of the transmitter (scatterer). $C_{A}$ and $C_{C}$ are constants, and $\tau_1$ and $\tau_2$ are the start and end times of the time window measured in lifetime units of the nuclear excited states, respectively.

We introduced a parameter ($\Delta w$), which indicates the center position of the absorption peak in the Mössbauer spectroscopy spectrum at the relative energy ($w_s$). $\Delta w$ is set to $0$ when generating the data in this study, and this means that the true value of the center position of the absorption peak in the spectrum is set to 0. Theoretically, if we set $\tau_1=0$ and $\tau_2= \infty$, then the left-hand side, $I(w_s,\Delta w, \tau_1, \tau_2)$ of Eq. (1) assumes a shape that can be approximated by a Lorentzian function ~\cite{seto2012condensed}. If the time window is chosen so that $\tau_2$ and $\tau_1$ are small, the signal intensity is relatively large because the time evolution shows the exponential decay, and the statistical preciseness of the spectrum can be easily improved. However, the line width increase, and the ambiguity of the absorption peak position in the spectrum is magnified. Conversely, if the time window is chosen so that $\tau_2$ and $\tau_1$ are large, the line width decreases and the ambiguity of the absorption peak position in the spectrum reduces, whereas the signal intensity decreases dramatically, making it difficult to improve the statistical precision of the spectrum. In addition, a small interval between $\tau_1$ and $\tau_2$ strengthens the effect of the limited time window, which generally makes the complex undulation in the background area more distinct.

Here, to reduce the computational cost associated with Eq. (1)- Eq. (3), the model is modified as in Eq. (6)-Eq. (8). The reason for modifying Eq. (1)- Eq. (3) as Eq. (6)- Eq. (8) is that Eq. (1) must be repeated when introducing the Bayesian estimation framework, and its computational cost is enormous.

\begin{widetext}
    \begin{equation}
    I\left(w_s,\Delta w, \tau\right)=I_{\mathrm{A}}\left(w_s,\Delta w, \tau\right)+I_{\mathrm{C}}\left(w_s,\Delta w, \tau\right)+I_{\mathrm{B}} \\
    \end{equation}
    \begin{equation}
    I_{\mathrm{A}}\left(w_s,\Delta w, \tau\right)=C_{\mathrm{A}} \int_0^{z_{\mathrm{s}}} \mathrm{d} z \left|\int \frac{\mathrm{d} w}{2 \pi} \frac{\exp (-\mathrm{i} w \tau)}{w-(w_{\mathrm{s}}+\Delta w)+\mathrm{i} / 2} E_{\mathrm{t}}(w) E_{\mathrm{s}}\left(w, w_{\mathrm{s}}+\Delta w, z\right)\right|^2 \
    \end{equation}
    \begin{equation}
    I_{\mathrm{C}}\left(w_s,\Delta w, \tau\right)=C_{\mathrm{C}} \int_0^{z_s} \mathrm{~d} z \left|\int \frac{\mathrm{d} w}{2 \pi} \exp (-\mathrm{i} w \tau)\left(E_{\mathrm{t}}(w) E_{\mathrm{s}}\left(w, w_{\mathrm{s}}+\Delta w, z\right)-1\right)\right|^2 \\
    \end{equation}
\end{widetext}

As shown in Eq. (7) and Eq. (8), it is possible to lower the computational cost by omitting the integral calculation of $\tau_1$ to $\tau_2$ in Eq. (2) and Eq. (3) and using one parameter, $\tau$, to represent time.

Moreover, $I\left(w_s,\Delta w,\tau_1,\tau_2\right)$ on the left-hand side of Eq. (1) can be calculated from $I\left(w_s,\Delta w,\tau_2\right)$ on the left  hand side of Eq. (6), as shown in Eq. (9):

\begin{equation}
    I(w_s,\Delta w,\tau_1,\tau_2) = \int_{\tau_1}^{\tau_2} \mathrm{~d} \tau I(w_s,\Delta w,\tau). 
\end{equation}

Next, in this study, Eq. (1) is defined as follow:
\begin{equation}
    f\left(w_s, \Theta \right) = I(w_s,\Delta w,\tau_1,\tau_2) = \int_{\tau_1}^{\tau_2} \mathrm{~d} \tau I(w_s,\Delta w,\tau), 
\end{equation}
where parameter $\Theta=\{\Delta w\}$ represents the parameter to be estimated in the resolution function. In this study, the simplest case is assumed, where the transmitter and scatterer are the same standard sample and conditions are the same as those in ~\cite{seto2010mossbauer}.

Since the observed data in a SR-based Mössbauer spectroscopy experiment indicate the count of the X-rays and electrons scattered from the scatterer, we assume that they are generated using a Poisson distribution. The observed data, $y$, is written as follows:

\begin{align}
    y &= \int_{\tau_1}^{\tau_2} \mathrm{~d} \tau \hspace{0.1cm} y_{\tau}\\
    y_{\tau} &\sim Poisson(I(w_s,\Delta w,\tau)) 
\end{align}

\subsection{Bayesian formulation}
In this section, we describe a framework for applying Bayesian estimation in SR-based Mössbauer spectroscopy.

In this study, data are generated for $\mathit{N}$ relative energies, ${w_s}^1,{w_s}^2, \cdots, {w_s}^N$. The generated observed data are $y_1, y_2, \cdots, y_N$. The generated observed data are modeled as a Poisson distribution, as shown in Eq. (12). The probability of obtaining observed data $y$ for a given relative energy $w_s$ is calculated as follows:
\begin{equation}
    P\left(y \mid w_s, \Theta\right)= \frac{f(w_s,\Theta)^y \exp(-f(w_s,\Theta))}{y!}
\end{equation}

Assuming that the observed dataset $Data = \{ {w_s}^i , y_i\}^{N}_{i=1}$,  follows an independent and identical distribution, the likelihood function is written as follows:
\begin{equation}
    P\left(Data \mid \Theta\right)= \prod_{i=1}^{N} P\left(y_{i} \mid {w_s}^i, \Theta\right) :=\exp (-N E_N(\Theta)).
\end{equation}

The cost function $E_N(\Theta)$ is defined as 
\begin{align}
    E_N(\Theta) =\frac{1}{N} \sum_{i=1}^{N}\left\{f({w_s}^i,\Theta)-y_{i} \log f({w_s}^i,\Theta) +\sum_{j=1}^{y_{i}} \log{j}\right\}.
\end{align}

From Eq. (14) of showing likelihood function, the posterior distribution, $P\left(\Theta \mid Data\right)$, can be written as follows using Bayes theorem:
\begin{align}
    P\left(\Theta \mid Data\right) &=\frac{P\left(Data \mid \Theta \right)P(\Theta)}{P(Data)}\\ 
        &=\frac{1}{Z} \exp (-N E_N(\Theta)) \rho(\Theta).
\end{align}
where $\mathit{Z}$ is the normalizing constant and $\rho(\Theta)$ is the prior distribution. In this study, the prior distribution is defined as $\rho(\Theta) = \rho(\Delta w)$. Thus, $\rho(\Delta w)$ is written as
\begin{align}
    \rho(\Delta w) = Uniform(\Delta w \mid \Delta w_{max}, \Delta w_{min}),
\end{align}
where $Uniform(\Delta w \mid \Delta w_{max}, \Delta w_{min})$ represents the uniform distribution of the maximum $\Delta w_{max}$ and minimum $\Delta w_{min}$.

\section{Numerical experiment}\label{sec:numerical}
The performance of the proposed method described in Section \ref{sec:formulation} is evaluated through numerical experiments using the generated data.

\begin{figure*}[th]
    \centering
    \includegraphics[width=100mm]{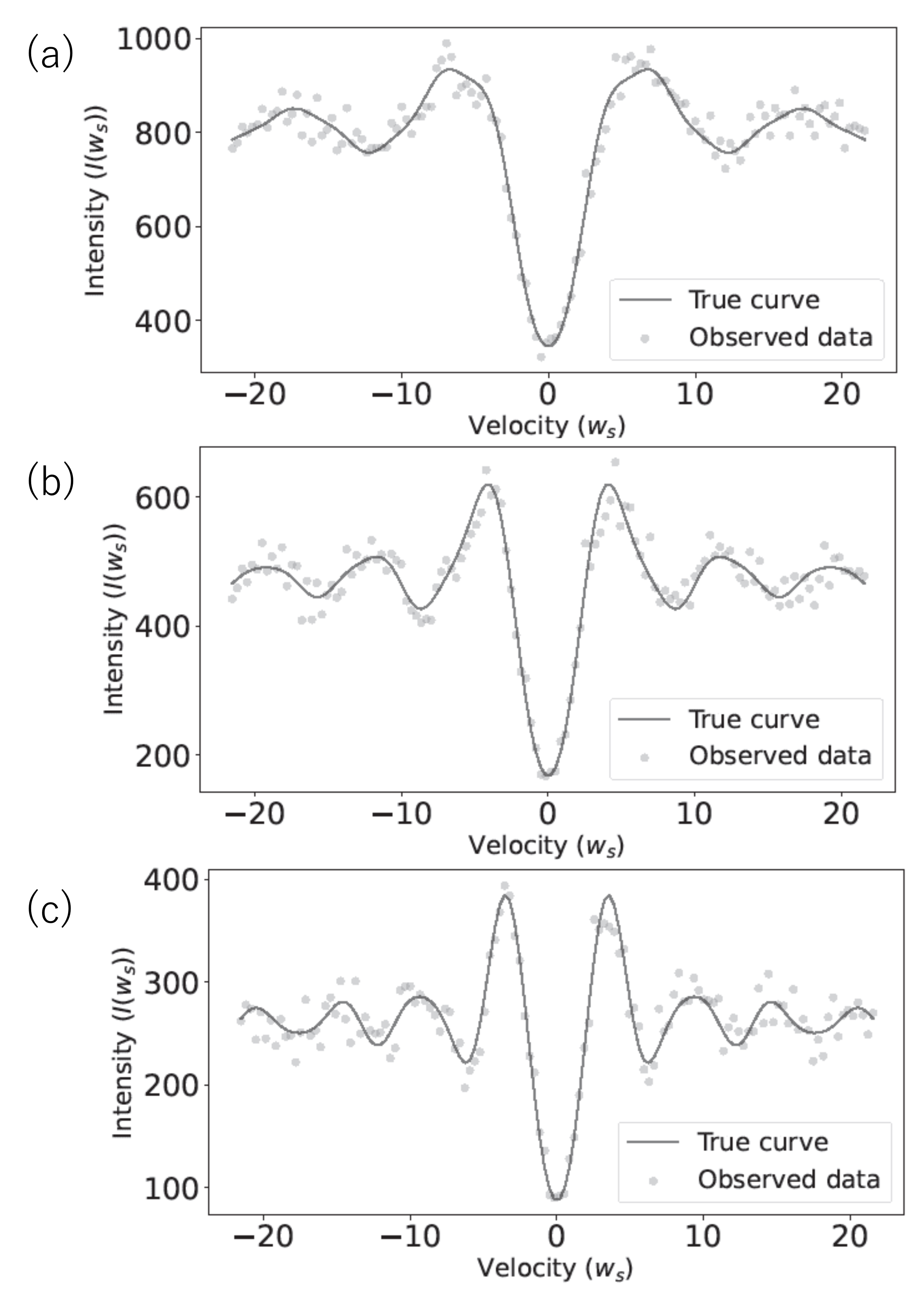}
    \caption{The SR-based Mössbauer spectroscopy data generated in the measurement windows of (a)$\tau_1=5.7 (ns)$, $\tau_2=17.0 (ns)$, (b)$\tau_1=8.1 (ns)$, $\tau_2=17.0 (ns)$, and (c)$\tau_1=10.5 (ns)$, $\tau_2=17.0 (ns)$. The horizontal axis expresses the relative velocity corresponding to the relative energy (mm/s), whereas the vertical axis express the intensity. The lines represents the $f(w_s,\Theta)$ values of the data generated from the resolution function, and the dots represents the values of the observed data, $Y$.}
\end{figure*}

\subsection{Generation of the observed data}\label{subsec:god}
When generating the observed dataset, $Data = \{ {w_s}^i , y_i\}^{N}_{i=1}$, in this study,  the variables that need to be specified are $\mathit{N}$, $\mathit{w_s}$, $\Delta w$, $\tau_1$, $\tau_2$, and $\mathit{M}$. As described, $\mathit{w_s}$ is the relative energy, $\Delta w$ is the center position of the absorption peak in the spectrum, and $\tau_1$ and $\tau_2$ are the start and end times of the measurement window, respectively. $\mathit{N}$ is the number of data, and $\mathit{M}$ is the discretization size of the measurement window. The discretization size ($\mathit{M}$) of the measurement window is a variable employed to numerically calculate the integral for $\tau$ in Eq. (10) and Eq. (11). In numerical calculations, Eq. (10) is calculated as Eq. (19) and Eq. (11) as Eq. (20):

\begin{equation}
    f\left(w_s, \Theta \right) = \sum_{\tau=\tau_1}^{\tau_2} I(w_s,\Delta w,\tau) 
\end{equation}

\begin{equation}
    y  = \sum_{\tau=\tau_1}^{\tau_2}  y_{\tau}\\
\end{equation}

The procedure for generating the observed data, $y$, is shown below. First, for a given $w$ and $\Theta = \{\Delta w\}$, the data from the resolution function are generated using Eq. (19). Next, we create the observed data ($y$) based on the assumption that the data from the resolution function follows a Poisson distribution, as in Eq. (20). We set $N=128$ and $\Delta w=0$. The range of $w_s$ is set to $-21.54 \sim 21.54$, and thus, ${w_s}^1=-21.54, {w_s}^2=-21.2007874,\cdots, {w_s}^{128}=21.54$.

Referring to \cite{seto2010mossbauer}, data with different $\tau_1$, $\tau_2$ values were generated for the three cases: (a), (b), and (c), as shown in Figure 1, and the values of $\tau_1$, $\tau_2$ in cases (a), (b), (c) are as follows:\\
(a)$\tau_1=5.7$, $\tau_2=17.0$, \\
(b)$\tau_1=8.1$, $\tau_2=17.0$, \\
(c)$\tau_1=10.5$, $\tau_2=17.0$\\

Further, note that $\mathit{M}=0.1$; therefore, the numbers of $\tau$ required to be added together in cases (a)–(c) are 114, 90, and 66, respectively. Figure 1 shows SR-based Mössbauer spectroscopy data generated for each case. The data are for the 151Eu Mössbauer spectroscopy spectra, where the lifetime of the excited level is 14 ns and the natural width is 1.3 mm/s. The horizontal axis represents the relative energy, whereas the vertical axis represents the intensity. The lines indicate the values of $f(w_s,\Theta)$, which are the data generated from the resolution functions, and the dots are the values of $y$, which are the generated observed data.

\begin{figure*}[th]
    \centering
    \includegraphics[width=179 mm]{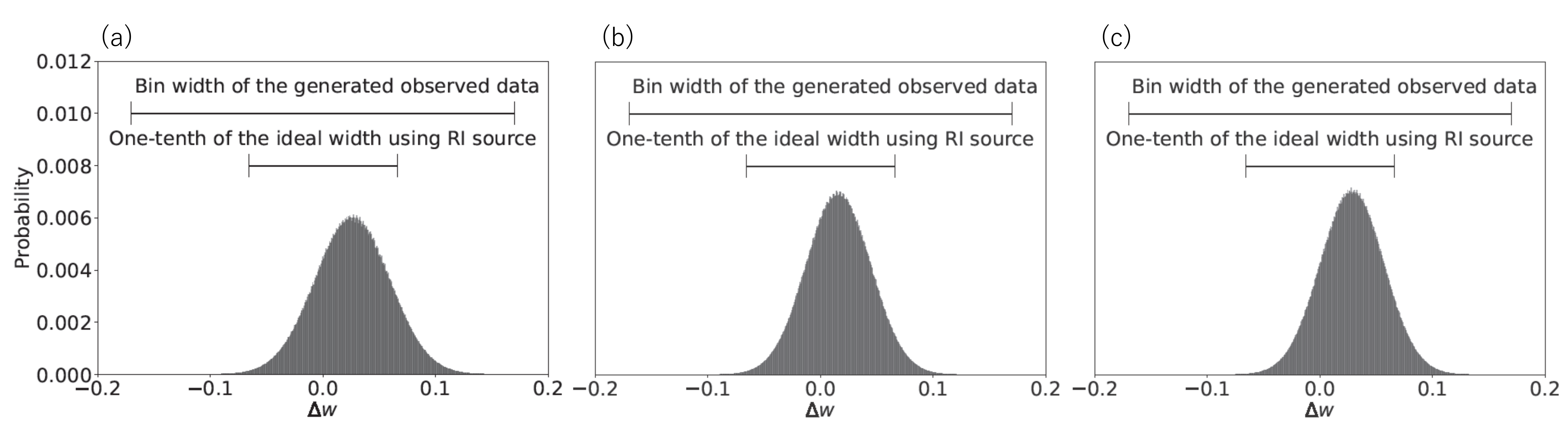}
    \caption{An example of the posterior distribution of the center of the spectrum ($\Delta w$), estimated by Bayesian estimation for the three cases of measurement windows(a)$\tau_1=5.7$, $\tau_2=17.0$, (b)$\tau_1=8.1$, $\tau_2=17.0$, (c)$\tau_1=10.5$, $\tau_2=17.0$. In the upper part of each figure, the bin width of the observed data in this study and the ideal line width in RI Mössbauer spectroscopy are shown. The horizontal axis represents the $\Delta w$ value, and the vertical axis represents the probability. The bars on the distribution are the typical sizes from the generated observed data. The upper bar is the bin width of the generated observed data, whereas the lower bar is one-tenth of the ideal experimental width in RI Mössbauer spectroscopy. Deviations that are reasonably less than these two typical sizes are experimentally negligible.}
\end{figure*}

\subsection{Calculation of the posterior distribution by Bayesian estimation}\label{subsec:cpd}
We describe the calculation of the posterior distribution of the peak center in the SR-based Mössbauer spectroscopy spectrum by Bayesian estimation.

First, the hyperparameters, $\Delta w_{max}$ and $\Delta w_{min}$, of the prior distribution ($\rho (\Theta)$) are set as $\Delta w_{max}=5$ and $\Delta w_{min}=-15$, respectively , based on previous research on 151Eu Mössbauer spectroscopy ~\cite{shenoy-wagner-1978}. Next, since the posterior distribution $P(\Theta \mid Data)$ only takes values above zero in the range of $\Delta w = -15 \sim \Delta w = 5$ for the setting of the prior distribution, the posterior distribution $P(\Theta \mid Data)$ can be estimated by calculating the cost function, $E_N(\Theta)$, in this range.

We calculated $E_N(\Delta)$ in 40,001 equal divisions in the range of $\Delta w=-15 \sim \Delta w=5$ and saved the results of $E_N(\Delta w=-15.0), E_N(\Delta w=-14.9995), E_N(\Delta w=-14.999),\cdots, E_N(\Delta w=5.0)$. Figure 2 shows an example of the posterior distribution calculated from the results of $E_N(\Theta)$. Figure 2 shows the bin width of the observed data in this study and one-tenth of the ideal line width using RI Mössbauer spectroscopy. No noticeable difference in the estimation performance is observed among measurement windows (a), (b), and (c) because the deviation of the estimated central position of the spectrum is less than one-fifth of the bin width of the observed data in this study.

\subsection{Evaluation of the posterior distribution}\label{subsec:eopd}
In this section, we first describe the evaluation method for the posterior distribution obtained from the calculations in Section \ref{subsec:cpd}. Thereafter, we compare the results with those of the conventional method. To compare the posterior distributions of $\Delta w$ for each case, we approximate the posterior distribution by a normal distribution and compare the mean and standard deviation of those normal distributions. This study follows the above procedure to generate data for 121 cases with $\tau_1$ given as 5.0 with 0.1 increments, and $\tau_2$ given as 17.0.

Figure 3 shows the error bar graphs resulting from changing the random seed of Poisson noise 10 times and approximating the posterior distribution for each case with a normal distribution. In Figure 3(a), the horizontal axis represents the value of  $\tau_1$ for each case, the vertical axis represents the mean of the approximated normal distribution, and the error bars indicate the standard error of the mean. In Figure 3(b), the horizontal axis represents $\tau_1$ for each case, the vertical axis represents the standard deviation of the approximated normal distribution, and the error bars represent the standard error of the standard deviation. The dashed lines in Figures 3(a) and 3(b) indicate the mean and standard deviation approximated by the normal distribution of the posterior distribution estimated by the conventional method using the Lorentzian function, respectively. Specifically, they are the mean and standard deviation of the posterior distribution estimated by the Lorentzian function for the spectral center position ($\Delta w$) for the observed data artificially generated by the resolution function, with $\tau_1=0, \tau_2=100$.

Considering the instrumental resolution obtained in actual SR-based Mössbauer spectroscopy experiments, no deviation is considered when the deviation from the peak center is extremely small. Intervals where the deviation is less than 0.015 and greater than -0.015 are indicated by dotted lines in Figure 3(a). Therefore, according to Figure 3(a), there is no deviation between the estimated position of the center of the spectrum and the true center in most cases using resolution function with the measurement window. In the interval, $\tau_1=[6, 11]$, the resolution function with a measurement window has a smaller deviation from the peak center than that for the conventional method using the Lorentzian function. Figure 3(b) shows that in the interval, $\tau_1=[6, 14]$, the standard deviation is relatively stable and small. As shown by the arrow in Figure 3(b), the $\tau_1$ with the smallest standard deviation is 9.8. In addition, the resolution function with the measurement window showed a standard deviation that was one-third smaller than that of the conventional method using the Lorentzian function. These results indicate that when the end time $\tau_2$ of the measurement window is fixed at 17.0, the estimation precision of the SR-based spectra is better than when the start time $\tau_1$ is in the $[6,14]$ interval. A comparison with the conventional method shows that the estimation precision of the resolution function using the proposed method is better than that achieved with the conventional method. In this study, we established a method that enables us to find the condition to determine resonance line position precisely by optimizing the measurement window.

\begin{figure*}[th]
    \centering
    \includegraphics[width=120mm]{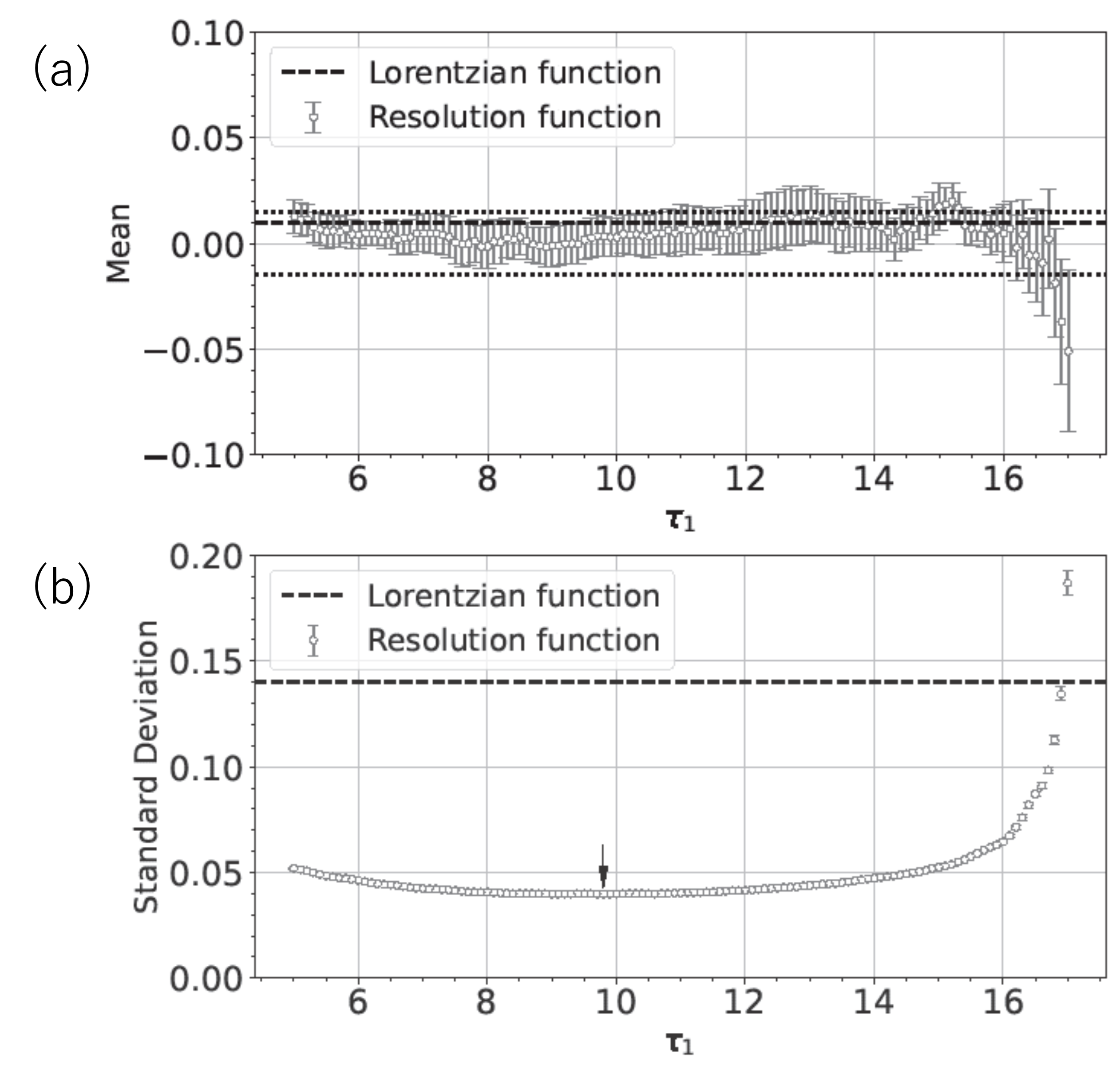}

    \caption{Error bar graphs resulting from changing the random seed of Poisson noise 10 times and approximating the posterior distribution of each case with a normal distribution. In (a), the horizontal axis represents the value of $\tau_1$ for each case, the vertical axis represents the mean of the approximated normal distribution, and the error bars indicate the standard errors of the mean. In (b), the horizontal axis represents the value of $\tau_1$ for each case, the vertical axis represents the standard deviation of the approximated normal distribution, and the error bars represent the standard error of the standard deviation. The dashed lines in (a) and (b) represent the mean and standard deviation approximated by the normal distribution of the posterior distribution estimated by the conventional method using the Lorentzian function, respectively. The dotted line in (a) indicates the interval at which the deviation is insignificant due to the instrumental resolution of the experiment. According to (a), based on actual experimental measurements, it can be assumed that there is a deviation less than the experimental limit between the estimated position of the center of the spectrum and the true center in most cases using resolution functions. In the interval, $\tau_1=[6, 11]$, the resolution function exhibits a smaller deviation from the peak center than the results observed for the conventional method using the Lorentzian function. According to (b), in the interval, $\tau_1=[6, 14]$, the standard deviation is relatively stable and small. As shown by the arrow in (b), the $\tau_1$ with the smallest standard deviation is 9.8. In addition, the performance of the resolution function with the measurement window is validated by the standard deviation being one-third smaller than that obtained using the conventional method, i.e., using the Lorentzian function. In this study, we established a method that enables us to find the condition to determine resonance line position precisely by optimizing the measurement window.
    }
\end{figure*}


\section{Summary}\label{sec:summary}
In this paper, we proposed a method to evaluate the position of absorption peaks in the SR-based Mössbauer spectroscopy spectra based on probability distributions by introducing Bayesian estimation to address the spectral resolution function of the measurement system considering the measurement window. The proposed method achieved the following:

\begin{enumerate}
    \item The estimation performances of theoretical models with different time windows were compared, and a theoretical background on the selection of the measurement window for SR-based Mössbauer spectroscopy was established.
    \item The theoretical model for determining the measurement window for SR-based Mössbauer spectroscopy is shown to be superior to the conventional simple fitting function method (i.e., the Lorentzian function).
    \item This method provides a theoretical basis for the analysis of general SR-based Mössbauer spectra in which a finite number of hyperfine interactions is observed using an optimized measurement window.
\end{enumerate}

Unlike the analysis of RI-based Mössbauer spectra in the previous study ~\cite{moriguchi2022bayesian}, the analysis of SR-based Mössbauer spectra generally requires computational resources for the optimization of the resolution function in addition to the selection of the Hamiltonian. However, we anticipate that the analysis approach of the SR-based Mössbauer spectra that involves observing a finite number of hyperfine interactions will become feasible, considering future computational resources. The knowledge obtained in this study will enable the precise analysis of hyperfine interactions in probe nuclei with more complex crystal and electronic structures by optimizing the measurement window for the SR-based Mössbauer spectrum analysis. The method established in this study is expected to serve as a foundation for further studies related to SR-based Mössbauer spectroscopy in materials science.

\begin{acknowledgments}
This work was supported by JST CREST (Grant Nos. PMJCR1761 and JPMJCR1861) from the Japan Science and Technology Agency (JST) and JSPS KAKENHI Grant-in-Aid for Scientific Research(A) (No. 23H00486).
\end{acknowledgments}


\bibliography{apssamp}

\begin{thebibliography}{20}%
\makeatletter
\providecommand \@ifxundefined [1]{%
 \@ifx{#1\undefined}
}%
\providecommand \@ifnum [1]{%
 \ifnum #1\expandafter \@firstoftwo
 \else \expandafter \@secondoftwo
 \fi
}%
\providecommand \@ifx [1]{%
 \ifx #1\expandafter \@firstoftwo
 \else \expandafter \@secondoftwo
 \fi
}%
\providecommand \natexlab [1]{#1}%
\providecommand \enquote  [1]{``#1''}%
\providecommand \bibnamefont  [1]{#1}%
\providecommand \bibfnamefont [1]{#1}%
\providecommand \citenamefont [1]{#1}%
\providecommand \href@noop [0]{\@secondoftwo}%
\providecommand \href [0]{\begingroup \@sanitize@url \@href}%
\providecommand \@href[1]{\@@startlink{#1}\@@href}%
\providecommand \@@href[1]{\endgroup#1\@@endlink}%
\providecommand \@sanitize@url [0]{\catcode `\\12\catcode `\$12\catcode `\&12\catcode `\#12\catcode `\^12\catcode `\_12\catcode `\%12\relax}%
\providecommand \@@startlink[1]{}%
\providecommand \@@endlink[0]{}%
\providecommand \url  [0]{\begingroup\@sanitize@url \@url }%
\providecommand \@url [1]{\endgroup\@href {#1}{\urlprefix }}%
\providecommand \urlprefix  [0]{URL }%
\providecommand \Eprint [0]{\href }%
\providecommand \doibase [0]{https://doi.org/}%
\providecommand \selectlanguage [0]{\@gobble}%
\providecommand \bibinfo  [0]{\@secondoftwo}%
\providecommand \bibfield  [0]{\@secondoftwo}%
\providecommand \translation [1]{[#1]}%
\providecommand \BibitemOpen [0]{}%
\providecommand \bibitemStop [0]{}%
\providecommand \bibitemNoStop [0]{.\EOS\space}%
\providecommand \EOS [0]{\spacefactor3000\relax}%
\providecommand \BibitemShut  [1]{\csname bibitem#1\endcsname}%
\let\auto@bib@innerbib\@empty
\bibitem [{\citenamefont {Zhdanov}\ and\ \citenamefont {Kuz'min}(1968)}]{zhdanov1968crystal}%
  \BibitemOpen
  \bibfield  {author} {\bibinfo {author} {\bibfnamefont {G.}~\bibnamefont {Zhdanov}}\ and\ \bibinfo {author} {\bibfnamefont {R.}~\bibnamefont {Kuz'min}},\ }\bibfield  {title} {\bibinfo {title} {Crystal structure investigations with the m{\"o}ssbauer effect},\ }\href@noop {} {\bibfield  {journal} {\bibinfo  {journal} {Acta Crystallographica Section B: Structural Crystallography and Crystal Chemistry}\ }\textbf {\bibinfo {volume} {24}},\ \bibinfo {pages} {10} (\bibinfo {year} {1968})}\BibitemShut {NoStop}%
\bibitem [{gue(2011)}]{guetlich2011}%
  \BibitemOpen
  \href@noop {} {\emph {\bibinfo {title} {M{\"o}ssbauer Spectroscopy and Transition Metal Chemistry}}}\ (\bibinfo  {publisher} {Springer},\ \bibinfo {year} {2011})\BibitemShut {NoStop}%
\bibitem [{\citenamefont {Seto}(2015)}]{seto2015}%
  \BibitemOpen
  \bibfield  {author} {\bibinfo {author} {\bibfnamefont {M.}~\bibnamefont {Seto}},\ }\bibfield  {title} {\bibinfo {title} {Project 13 project research on the advanced utilization of multi-element mössbauer spectroscopy for the study of condensed matter},\ }\href@noop {} {\bibfield  {journal} {\bibinfo  {journal} {KURRI Progress Report}\ }\textbf {\bibinfo {volume} {2014 (APRIL 2014 – MARCH 2015)}},\ \bibinfo {pages} {75} (\bibinfo {year} {2015})}\BibitemShut {NoStop}%
\bibitem [{\citenamefont {Seto}\ \emph {et~al.}(2009)\citenamefont {Seto}, \citenamefont {Masuda}, \citenamefont {Higashitaniguchi}, \citenamefont {Kitao}, \citenamefont {Kobayashi}, \citenamefont {Inaba}, \citenamefont {Mitsui},\ and\ \citenamefont {Yoda}}]{seto2009synchrotron}%
  \BibitemOpen
  \bibfield  {author} {\bibinfo {author} {\bibfnamefont {M.}~\bibnamefont {Seto}}, \bibinfo {author} {\bibfnamefont {R.}~\bibnamefont {Masuda}}, \bibinfo {author} {\bibfnamefont {S.}~\bibnamefont {Higashitaniguchi}}, \bibinfo {author} {\bibfnamefont {S.}~\bibnamefont {Kitao}}, \bibinfo {author} {\bibfnamefont {Y.}~\bibnamefont {Kobayashi}}, \bibinfo {author} {\bibfnamefont {C.}~\bibnamefont {Inaba}}, \bibinfo {author} {\bibfnamefont {T.}~\bibnamefont {Mitsui}},\ and\ \bibinfo {author} {\bibfnamefont {Y.}~\bibnamefont {Yoda}},\ }\bibfield  {title} {\bibinfo {title} {Synchrotron-radiation-based m{\"o}ssbauer spectroscopy},\ }\href@noop {} {\bibfield  {journal} {\bibinfo  {journal} {Physical review letters}\ }\textbf {\bibinfo {volume} {102}},\ \bibinfo {pages} {217602} (\bibinfo {year} {2009})}\BibitemShut {NoStop}%
\bibitem [{\citenamefont {Seto}\ \emph {et~al.}(2010)\citenamefont {Seto}, \citenamefont {Masuda}, \citenamefont {Higashitaniguchi}, \citenamefont {Kitao}, \citenamefont {Kobayashi}, \citenamefont {Inaba}, \citenamefont {Mitsui},\ and\ \citenamefont {Yoda}}]{seto2010mossbauer}%
  \BibitemOpen
  \bibfield  {author} {\bibinfo {author} {\bibfnamefont {M.}~\bibnamefont {Seto}}, \bibinfo {author} {\bibfnamefont {R.}~\bibnamefont {Masuda}}, \bibinfo {author} {\bibfnamefont {S.}~\bibnamefont {Higashitaniguchi}}, \bibinfo {author} {\bibfnamefont {S.}~\bibnamefont {Kitao}}, \bibinfo {author} {\bibfnamefont {Y.}~\bibnamefont {Kobayashi}}, \bibinfo {author} {\bibfnamefont {C.}~\bibnamefont {Inaba}}, \bibinfo {author} {\bibfnamefont {T.}~\bibnamefont {Mitsui}},\ and\ \bibinfo {author} {\bibfnamefont {Y.}~\bibnamefont {Yoda}},\ }\bibfield  {title} {\bibinfo {title} {M{\"o}ssbauer spectroscopy in the energy domain using synchrotron radiation},\ }in\ \href@noop {} {\emph {\bibinfo {booktitle} {Journal of Physics: Conference Series}}},\ Vol.\ \bibinfo {volume} {217}\ (\bibinfo {organization} {IOP Publishing},\ \bibinfo {year} {2010})\ p.\ \bibinfo {pages} {012002}\BibitemShut {NoStop}%
\bibitem [{\citenamefont {Seto}\ \emph {et~al.}(2021)\citenamefont {Seto}, \citenamefont {Masuda},\ and\ \citenamefont {Saito}}]{seto2021synchrotron}%
  \BibitemOpen
  \bibfield  {author} {\bibinfo {author} {\bibfnamefont {M.}~\bibnamefont {Seto}}, \bibinfo {author} {\bibfnamefont {R.}~\bibnamefont {Masuda}},\ and\ \bibinfo {author} {\bibfnamefont {M.}~\bibnamefont {Saito}},\ }\bibinfo {title} {Synchrotron-radiation-based energy-domain mössbauer spectroscopy, nuclear resonant inelastic scattering, and quasielastic scattering using mössbauer gamma rays}\ (\bibinfo  {publisher} {Springer Singapore},\ \bibinfo {address} {Singapore},\ \bibinfo {year} {2021})\BibitemShut {NoStop}%
\bibitem [{\citenamefont {Masuda}\ \emph {et~al.}(2014)\citenamefont {Masuda}, \citenamefont {Kobayashi}, \citenamefont {Kitao}, \citenamefont {Kurokuzu}, \citenamefont {Saito}, \citenamefont {Yoda}, \citenamefont {Mitsui}, \citenamefont {Iga},\ and\ \citenamefont {Seto}}]{masuda2014synchrotron}%
  \BibitemOpen
  \bibfield  {author} {\bibinfo {author} {\bibfnamefont {R.}~\bibnamefont {Masuda}}, \bibinfo {author} {\bibfnamefont {Y.}~\bibnamefont {Kobayashi}}, \bibinfo {author} {\bibfnamefont {S.}~\bibnamefont {Kitao}}, \bibinfo {author} {\bibfnamefont {M.}~\bibnamefont {Kurokuzu}}, \bibinfo {author} {\bibfnamefont {M.}~\bibnamefont {Saito}}, \bibinfo {author} {\bibfnamefont {Y.}~\bibnamefont {Yoda}}, \bibinfo {author} {\bibfnamefont {T.}~\bibnamefont {Mitsui}}, \bibinfo {author} {\bibfnamefont {F.}~\bibnamefont {Iga}},\ and\ \bibinfo {author} {\bibfnamefont {M.}~\bibnamefont {Seto}},\ }\bibfield  {title} {\bibinfo {title} {Synchrotron radiation-based mössbauer spectra of \(^{174}\)yb measured with internal conversion electrons},\ }\href@noop {} {\bibfield  {journal} {\bibinfo  {journal} {Applied Physics Letters}\ }\textbf {\bibinfo {volume} {104}},\ \bibinfo {pages} {082411} (\bibinfo {year} {2014})}\BibitemShut {NoStop}%
\bibitem [{\citenamefont {seto}(2012)}]{seto2012condensed}%
  \BibitemOpen
  \bibfield  {author} {\bibinfo {author} {\bibfnamefont {M.}~\bibnamefont {seto}},\ }\bibfield  {title} {\bibinfo {title} {Condensed matter physics using nuclear resonant scattering},\ }\href@noop {} {\bibfield  {journal} {\bibinfo  {journal} {Journal of the Physical Society of Japan}\ }\textbf {\bibinfo {volume} {82}},\ \bibinfo {pages} {021016} (\bibinfo {year} {2012})}\BibitemShut {NoStop}%
\bibitem [{\citenamefont {Feynman}(2011)}]{matsuoka2011structural}%
  \BibitemOpen
  \bibfield  {author} {\bibinfo {author} {\bibfnamefont {R.~P.}\ \bibnamefont {Feynman}},\ }\bibfield  {title} {\bibinfo {title} {Structural and valence changes of europium hydride induced by application of high-pressure \(h_2\)},\ }\href@noop {} {\bibfield  {journal} {\bibinfo  {journal} {Physical Review Letters}\ }\textbf {\bibinfo {volume} {107}},\ \bibinfo {pages} {025501} (\bibinfo {year} {2011})}\BibitemShut {NoStop}%
\bibitem [{\citenamefont {Tsutsui}\ \emph {et~al.}(2018)\citenamefont {Tsutsui}, \citenamefont {Masuda}, \citenamefont {Yoda},\ and\ \citenamefont {Seto}}]{tsutsui2018precise}%
  \BibitemOpen
  \bibfield  {author} {\bibinfo {author} {\bibfnamefont {S.}~\bibnamefont {Tsutsui}}, \bibinfo {author} {\bibfnamefont {R.}~\bibnamefont {Masuda}}, \bibinfo {author} {\bibfnamefont {Y.}~\bibnamefont {Yoda}},\ and\ \bibinfo {author} {\bibfnamefont {M.}~\bibnamefont {Seto}},\ }\bibfield  {title} {\bibinfo {title} {Precise determination of hyperfine interactions and second-order doppler shift in \(^{149}\)sm mössbauer transition},\ }\href@noop {} {\bibfield  {journal} {\bibinfo  {journal} {Hyperfine Interactions}\ }\textbf {\bibinfo {volume} {239}},\ \bibinfo {pages} {50} (\bibinfo {year} {2018})}\BibitemShut {NoStop}%
\bibitem [{\citenamefont {Akai}\ \emph {et~al.}(2018)\citenamefont {Akai}, \citenamefont {Iwamitsu},\ and\ \citenamefont {Okada}}]{Akai_2018}%
  \BibitemOpen
  \bibfield  {author} {\bibinfo {author} {\bibfnamefont {I.}~\bibnamefont {Akai}}, \bibinfo {author} {\bibfnamefont {K.}~\bibnamefont {Iwamitsu}},\ and\ \bibinfo {author} {\bibfnamefont {M.}~\bibnamefont {Okada}},\ }\bibfield  {title} {\bibinfo {title} {Bayesian spectroscopy in solid-state photo-physics},\ }\href@noop {} {\bibfield  {journal} {\bibinfo  {journal} {Journal of Physics: Conference Series}\ }\textbf {\bibinfo {volume} {1036}},\ \bibinfo {pages} {012022} (\bibinfo {year} {2018})}\BibitemShut {NoStop}%
\bibitem [{\citenamefont {Mototake}\ \emph {et~al.}(2019)\citenamefont {Mototake}, \citenamefont {Mizumaki}, \citenamefont {Akai},\ and\ \citenamefont {Okada}}]{Mototake2019}%
  \BibitemOpen
  \bibfield  {author} {\bibinfo {author} {\bibfnamefont {Y.}~\bibnamefont {Mototake}}, \bibinfo {author} {\bibfnamefont {M.}~\bibnamefont {Mizumaki}}, \bibinfo {author} {\bibfnamefont {I.}~\bibnamefont {Akai}},\ and\ \bibinfo {author} {\bibfnamefont {M.}~\bibnamefont {Okada}},\ }\bibfield  {title} {\bibinfo {title} {Bayesian hamiltonian selection in x-ray photoelectron spectroscopy},\ }\href@noop {} {\bibfield  {journal} {\bibinfo  {journal} {Journal of the Physical Society of Japan}\ }\textbf {\bibinfo {volume} {88}},\ \bibinfo {pages} {034004} (\bibinfo {year} {2019})}\BibitemShut {NoStop}%
\bibitem [{\citenamefont {Kashiwamura}\ \emph {et~al.}(2022)\citenamefont {Kashiwamura}, \citenamefont {Katakami}, \citenamefont {Yamagami}, \citenamefont {Iwamitsu}, \citenamefont {Kumazoe}, \citenamefont {Nagata}, \citenamefont {Okajima}, \citenamefont {Akai},\ and\ \citenamefont {Okada}}]{Kashiwamura2022}%
  \BibitemOpen
  \bibfield  {author} {\bibinfo {author} {\bibfnamefont {S.}~\bibnamefont {Kashiwamura}}, \bibinfo {author} {\bibfnamefont {S.}~\bibnamefont {Katakami}}, \bibinfo {author} {\bibfnamefont {R.}~\bibnamefont {Yamagami}}, \bibinfo {author} {\bibfnamefont {K.}~\bibnamefont {Iwamitsu}}, \bibinfo {author} {\bibfnamefont {H.}~\bibnamefont {Kumazoe}}, \bibinfo {author} {\bibfnamefont {K.}~\bibnamefont {Nagata}}, \bibinfo {author} {\bibfnamefont {T.}~\bibnamefont {Okajima}}, \bibinfo {author} {\bibfnamefont {I.}~\bibnamefont {Akai}},\ and\ \bibinfo {author} {\bibfnamefont {M.}~\bibnamefont {Okada}},\ }\bibfield  {title} {\bibinfo {title} {Bayesian spectral deconvolution of x-ray absorption near edge structure discriminating between high- and low-energy domains},\ }\href@noop {} {\bibfield  {journal} {\bibinfo  {journal} {Journal of the Physical Society of Japan}\ }\textbf {\bibinfo {volume} {91}},\ \bibinfo {pages} {074009} (\bibinfo {year} {2022})}\BibitemShut {NoStop}%
\bibitem [{\citenamefont {Ueda}\ \emph {et~al.}(2023)\citenamefont {Ueda}, \citenamefont {Katakami}, \citenamefont {Yoshida}, \citenamefont {Koyama}, \citenamefont {Nakai}, \citenamefont {Mito}, \citenamefont {Mizumaki},\ and\ \citenamefont {Okada}}]{Ueda2023}%
  \BibitemOpen
  \bibfield  {author} {\bibinfo {author} {\bibfnamefont {H.}~\bibnamefont {Ueda}}, \bibinfo {author} {\bibfnamefont {S.}~\bibnamefont {Katakami}}, \bibinfo {author} {\bibfnamefont {S.}~\bibnamefont {Yoshida}}, \bibinfo {author} {\bibfnamefont {T.}~\bibnamefont {Koyama}}, \bibinfo {author} {\bibfnamefont {Y.}~\bibnamefont {Nakai}}, \bibinfo {author} {\bibfnamefont {T.}~\bibnamefont {Mito}}, \bibinfo {author} {\bibfnamefont {M.}~\bibnamefont {Mizumaki}},\ and\ \bibinfo {author} {\bibfnamefont {M.}~\bibnamefont {Okada}},\ }\bibfield  {title} {\bibinfo {title} {Bayesian approach to t1 analysis in nmr spectroscopy with applications to solid state physics},\ }\href@noop {} {\bibfield  {journal} {\bibinfo  {journal} {Journal of the Physical Society of Japan}\ }\textbf {\bibinfo {volume} {92}},\ \bibinfo {pages} {054002} (\bibinfo {year} {2023})}\BibitemShut {NoStop}%
\bibitem [{\citenamefont {Moriguchi}\ \emph {et~al.}(2022)\citenamefont {Moriguchi}, \citenamefont {Tsutsui}, \citenamefont {Katakami}, \citenamefont {Nagata}, \citenamefont {Mizumaki},\ and\ \citenamefont {Okada}}]{moriguchi2022bayesian}%
  \BibitemOpen
  \bibfield  {author} {\bibinfo {author} {\bibfnamefont {R.}~\bibnamefont {Moriguchi}}, \bibinfo {author} {\bibfnamefont {S.}~\bibnamefont {Tsutsui}}, \bibinfo {author} {\bibfnamefont {S.}~\bibnamefont {Katakami}}, \bibinfo {author} {\bibfnamefont {K.}~\bibnamefont {Nagata}}, \bibinfo {author} {\bibfnamefont {M.}~\bibnamefont {Mizumaki}},\ and\ \bibinfo {author} {\bibfnamefont {M.}~\bibnamefont {Okada}},\ }\bibfield  {title} {\bibinfo {title} {Bayesian inference on hamiltonian selections for m{\"o}ssbauer spectroscopy},\ }\href@noop {} {\bibfield  {journal} {\bibinfo  {journal} {Journal of the Physical Society of Japan}\ }\textbf {\bibinfo {volume} {91}},\ \bibinfo {pages} {104002} (\bibinfo {year} {2022})}\BibitemShut {NoStop}%
\bibitem [{\citenamefont {Toussaint}(2011)}]{RevModPhys.83.943}%
  \BibitemOpen
  \bibfield  {author} {\bibinfo {author} {\bibfnamefont {U.}~\bibnamefont {Toussaint}},\ }\bibfield  {title} {\bibinfo {title} {Bayesian inference in physics},\ }\href@noop {} {\bibfield  {journal} {\bibinfo  {journal} {Rev. Mod. Phys.}\ }\textbf {\bibinfo {volume} {83}},\ \bibinfo {pages} {943} (\bibinfo {year} {2011})}\BibitemShut {NoStop}%
\bibitem [{\citenamefont {Nagata}\ \emph {et~al.}(2012)\citenamefont {Nagata}, \citenamefont {Sugita},\ and\ \citenamefont {Okada}}]{nagata2012bayesian}%
  \BibitemOpen
  \bibfield  {author} {\bibinfo {author} {\bibfnamefont {K.}~\bibnamefont {Nagata}}, \bibinfo {author} {\bibfnamefont {S.}~\bibnamefont {Sugita}},\ and\ \bibinfo {author} {\bibfnamefont {M.}~\bibnamefont {Okada}},\ }\bibfield  {title} {\bibinfo {title} {Bayesian spectral deconvolution with the exchange monte carlo method},\ }\href@noop {} {\bibfield  {journal} {\bibinfo  {journal} {Neural Networks}\ }\textbf {\bibinfo {volume} {28}},\ \bibinfo {pages} {82} (\bibinfo {year} {2012})}\BibitemShut {NoStop}%
\bibitem [{\citenamefont {Tokuda}\ \emph {et~al.}(2017)\citenamefont {Tokuda}, \citenamefont {Nagata},\ and\ \citenamefont {Okada}}]{Tokuda2017}%
  \BibitemOpen
  \bibfield  {author} {\bibinfo {author} {\bibfnamefont {S.}~\bibnamefont {Tokuda}}, \bibinfo {author} {\bibfnamefont {K.}~\bibnamefont {Nagata}},\ and\ \bibinfo {author} {\bibfnamefont {M.}~\bibnamefont {Okada}},\ }\bibfield  {title} {\bibinfo {title} {Simultaneous estimation of noise variance and number of peaks in bayesian spectral deconvolution},\ }\href@noop {} {\bibfield  {journal} {\bibinfo  {journal} {Journal of the Physical Society of Japan}\ }\textbf {\bibinfo {volume} {86}},\ \bibinfo {pages} {024001} (\bibinfo {year} {2017})}\BibitemShut {NoStop}%
\bibitem [{\citenamefont {Nagata}\ \emph {et~al.}(2019)\citenamefont {Nagata}, \citenamefont {Muraoka}, \citenamefont {Mototake}, \citenamefont {Sasaki},\ and\ \citenamefont {Okada}}]{nagata2019bayesian}%
  \BibitemOpen
  \bibfield  {author} {\bibinfo {author} {\bibfnamefont {K.}~\bibnamefont {Nagata}}, \bibinfo {author} {\bibfnamefont {R.}~\bibnamefont {Muraoka}}, \bibinfo {author} {\bibfnamefont {Y.}~\bibnamefont {Mototake}}, \bibinfo {author} {\bibfnamefont {T.}~\bibnamefont {Sasaki}},\ and\ \bibinfo {author} {\bibfnamefont {M.}~\bibnamefont {Okada}},\ }\bibfield  {title} {\bibinfo {title} {Bayesian spectral deconvolution based on poisson distribution: Bayesian measurement and virtual measurement analytics (vma)},\ }\href@noop {} {\bibfield  {journal} {\bibinfo  {journal} {Journal of the Physical Society of Japan}\ }\textbf {\bibinfo {volume} {88}},\ \bibinfo {pages} {044003} (\bibinfo {year} {2019})}\BibitemShut {NoStop}%
\bibitem [{\citenamefont {Shenoy}\ and\ \citenamefont {Wagner}(1978)}]{shenoy-wagner-1978}%
  \BibitemOpen
  \bibinfo {editor} {\bibfnamefont {G.}~\bibnamefont {Shenoy}}\ and\ \bibinfo {editor} {\bibfnamefont {F.}~\bibnamefont {Wagner}},\ eds.,\ \href@noop {} {\emph {\bibinfo {title} {M{\"o}ssbauer Isomer Shifts}}}\ (\bibinfo  {publisher} {North-Holland},\ \bibinfo {year} {1978})\BibitemShut {NoStop}%
\end{thebibliography}%

\end{document}